\documentclass{ptapap}
\usepackage{amsmath}
\usepackage{amssymb}

\author{Jadwiga Daszy\'nska-Daszkiewicz}[UWR]
\author{Przemys{\l}aw Walczak}[UWR]
\author{Alosha Pamyatnykh}[CAMK]
\author{Wojciech Szewczuk}[UWR]
\affil[UWR]{Astronomical Institute, Wroc{\l}aw University, ul. Kopernika 11, 51--622 Wroc{\l}aw, Poland}
\affil[CAMK]{Nicolaus Copernicus Astronomical Center, Polish Academy of Sciences, Bartycka 18, 00--716 Warsaw, Poland}

\title{Testing stellar opacities using asteroseismology}

\begin{document}

\maketitle

\begin{abstract}

We present what constraints on opacities can be derived from the analysis of stellar pulsations of BA-type main-sequence stars.
This analysis consists of the construction of complex seismic models which reproduce the observed frequencies
as well as the bolometric flux amplitude extracted from the multi-colour photometric variations.
Stellar seismology, i.e., {\it asteroseismology}, is a relatively young branch of astrophysics
and, currently, provides the most accurate test of the theory of internal structure and evolution.
We show that opacities under stellar conditions need to be modified at the depth of temperatures
$T=110~000-290~000$\,K. The revision of opacity data is of great importance because they are crucial
for all branches of astrophysics.

\end{abstract}

\section{Introduction}

Opacity data are one of the main ingredients in stellar modelling. They
determine the transport of energy and, consequently, the internal structure of a star.
The values of opacities define also preconditions for excitation of heat-driven
pulsations as observed, e.g., in main-sequence or classical variables.

For many years, incorrect opacity data were used in astrophysics, before they were recomputed in the early nineties
by the two independent teams: OPAL \citep{Iglesias1992, Rogers1992} and OP \citep{Seaton1993, Seaton1994}.
The most spectacular result was the finding of a local maximum caused by a huge number of transition lines of iron group elements.
This new bump occurs at temperature of about 200~000 K and is called the Z-bump.

The discovery of the Z-bump was a big step forward in stellar physics, however there are still some uncertainties
and many indications that something is still missing and/or has not been correctly included. As a consequence,
stellar opacities can be still underestimated or overestimated at some temperatures \citep{Blancard2016}.

The first example is the problem with the modelling of the Sun, a fundamental calibrator of stellar structure and evolution.
The revision of solar chemical abundances \citep{Asplund2005, Asplund2009} caused a disagreement between the standard solar model
and the helioseismic and neutrino-flux predictions \citep[e.g.,][]{Turck-Chieze2004, Guzik2008}.
An increase of the opacity  in the solar radiative zone would solve the problem
\citep[e.g.,][]{Christensen-Dalsgaard2009}.
Indeed, the laboratory measurements in physical conditions similar to the boundary of the solar convection zone
have indicated that the Rosseland mean opacities of iron predicted by all codes are underestimated by 30 to 400 \%
\citep{Bailey2015, Pradhan2018, Zhao2018}.
However, these increases of opacity are not sufficient to solve the problem and other drawbacks in the solar modelling have to be identified \citep{Iglesias2017}.

The other example indicating a possible problem with the opacity data are the main-sequence pulsators of B and A spectral type
that simultaneously exhibit both pressure  and buoyancy (gravity) modes.
So far, none of pulsational models computed  with the standard opacity tables can account for these hybrid pulsations \citep[e.g.,][]{Pamyatnykh2004, JDD2017, Balona2015}.
Here, we summarize our results on constraints on opacities derived from seismic studies of a few B- and A-type main sequence pulsators.

\section{Stellar opacities}

The values of opacity represent ability of stellar material to absorb radiation.
To compute the opacity data all microscopic processes in the plasma at each photon
frequency $\nu$ need to be considered, i.e., bound-bound, bound-free, free-free and electron scattering processes.
At lower temperatures, one has to take into account opacity sources arising from the negative hydrogen ion H$^{-}$,
molecules (e.g., TiO, CO, C$_2$, CN, C$_2$H$_2$) and even dust grains.
In case of degenerate matter, a ,,conductive opacity'' needs to be considered.

The monochromatic opacity coefficient, $\kappa_\nu$, is expressed as
\begin{equation}
\kappa_\nu=\sigma_\nu\frac{N}{\rho} ~~~  [{\rm cm}^2 {\rm g}^{-1}] \label{eq1}
\end{equation}
where $\sigma$ is the cross-section for an interaction, $\rho$ is the density and $N$ is the number of particles per volume unit.
In other words, the opacity coefficient, $\kappa_\nu$ is the effective cross section per unit mass.
As a consequence, the value of $\kappa_\nu$ defines the mean free path for a photon of the energy $E=h\nu$
and appears in the equation of energy transport.

Computations of opacity data is not a trivial task because they demand complicated and laborious calculations of atomic data,
that is: ground and excited energy levels, oscillator strengths and photoionization cross sections for all considered elements.
Besides an adequate equation of state is needed in a wide range of temperatures and densities.
Therefore, the values of opacities do not depend only on the abundance of individual
elements but also on how they are computed, in particular on the number of fine structure energy levels taken into account for each ion.

There are three main databases of the opacity tables widely used for evolutionary modelling of normal (main sequence) stars:
OPAL \citep{Iglesias1996}, OP \citep{Seaton1996, Seaton2005} and OPLIB \citep{Colgan2015, Colgan2016}.
At temperatures below 9000\,K, these tables are supplemented by data of \citet{Ferguson2005}.

There are still some differences between opacities from these databases  resulting from the adopted physics and methods of computations.
In Fig.\,1, we show a comparison of OPAL, OP and OPLIB opacities for two models of masses 10 and 1.8$M_\odot$ and effective temperatures $\log T_{\rm eff}=4.35$ and 3.90,  respectively.
In the top panels, we show the run of the mean Rosseland opacity inside these models and in the bottom panels
the corresponding values of the temperature derivative $\kappa_T=\partial\log\kappa/\partial\log T$.
    \begin{figure}
    \includegraphics[width=\textwidth, height=4.8cm]{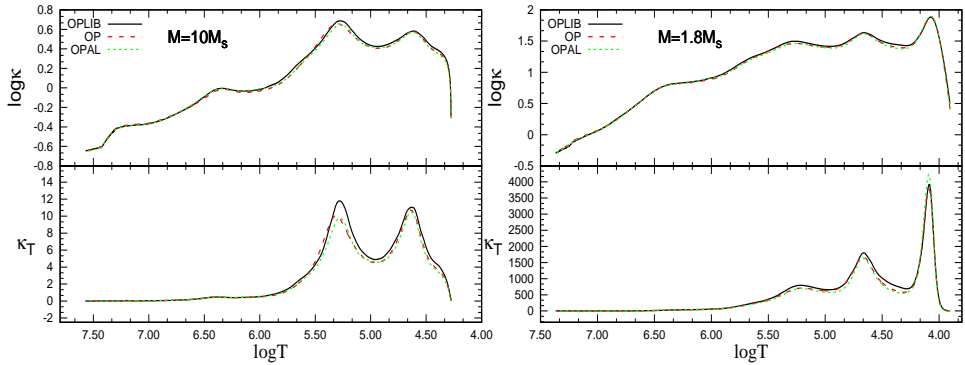}
    \caption{The run of the mean Rosseland opacity (the top panels) inside the models of masses 10 and 1.8$M_\odot$ and effective temperatures 
    $\log T_{\rm eff}=4.35$ and 3.90, respectively. The corresponding temperature derivatives of $\kappa$ are plotted in the bottom panels.}
    \label{fig:1}
    \end{figure}

\section{Seismic modelling of $\beta$ Cep/SPB and $\delta$ Sct stars}

Pulsating stars are intrinsic variables that change their brightness  and/or line profiles due to physical changes within their interiors,
namely because of propagation of hydrodynamic waves.
Stars can pulsate in many modes which penetrate different parts of a star and have different sensitivities to its structure.
By observing multiple modes, one can therefore infer information on the internal structure and dynamic.

To this end, a grid of seismic models needs to be constructed. Such models have eigenfrequencies  which reproduce
within the observational errors the observed frequencies. The theoretical frequency
of a given mode depends on parameters of the model, like a mass, age, chemical composition, angular momentum,
and free parameters of theory describing such phenomena as convection, overshooting from convective regions,
mass loss, angular momentum evolution, mixing processes etc.

Because we analyse the heat-driven pulsators the next requirement in addition to frequency matching is mode excitation,
that is a good seismic model should have eigenmodes corresponding to the observed frequencies unstable.
In pulsational modelling of BA-type main sequence stars this is a key problem because any standard-opacity model cannot account
for the whole oscillation spectra.

The old classification, mainly before the era of space missions, distinguished two types of B-type main sequence pulsators: $\beta$ Cephei  stars with B0-B2.5 spectral types
and Slowly Pulsating B-type (SPB) stars with B3-B8 spectral types.
Pulsations in both these types are driven by the opacity mechanism acting at the Z-bump ($T\approx 200~000$\,K),
but in the first ones pressure modes of high frequencies were observed and in the latter -- high-order gravity modes of low frequencies.
In those days, these observational facts perfectly agreed with predictions from theory of pulsation.

Observation from space by MOST, CoRoT, Kepler, BRITE decreased enormously the detection level  of the pulsational amplitudes
and revealed high-order gravity modes in early B-type pulsators and pressure modes in late B-type pulsators.
Likewise, in particular the Kepler photometry revealed that all $\delta$ Scuti stars show both high
and low frequencies \citep[e.g.,][]{Balona2014}.
These low frequencies can only be understood in terms of pulsation in high-order gravity modes.
$\delta$ Scuti stars pulsate due to $\kappa$ mechanism operating in the HeII ionization zone and were previously
thought as pulsating stars only  in pressure and mixed modes.

The problem with mode excitation is presented in Fig.\,2 where the normalized instability parameter $\eta$ is plotted as a function of the mode frequency for the degrees $\ell=0,1$ and 2.
The left panel corresponds to the seismic models of early B-type pulsator $\nu$ Eridani computed with the three sources of opacity data; OPAL, OP and OPLIB.
The oscillation spectrum is from the BRITE photometry with the amplitude values on the right Y-axis.
The right panel shows the run of $\eta(\nu)$ for the $\delta$ Sct model suitable for the star KIC\,8197788.
As one can see only modes in the p-mode frequency range are excited with an exception of a few quadrupole modes in case of $\beta$ Cep model
computed with the OP tables. However, these few modes cannot explain all low frequency peaks observed in many (if not all) early B-type pulsators.
\begin{figure*}
	\centering
\includegraphics[clip,width=0.48\linewidth]{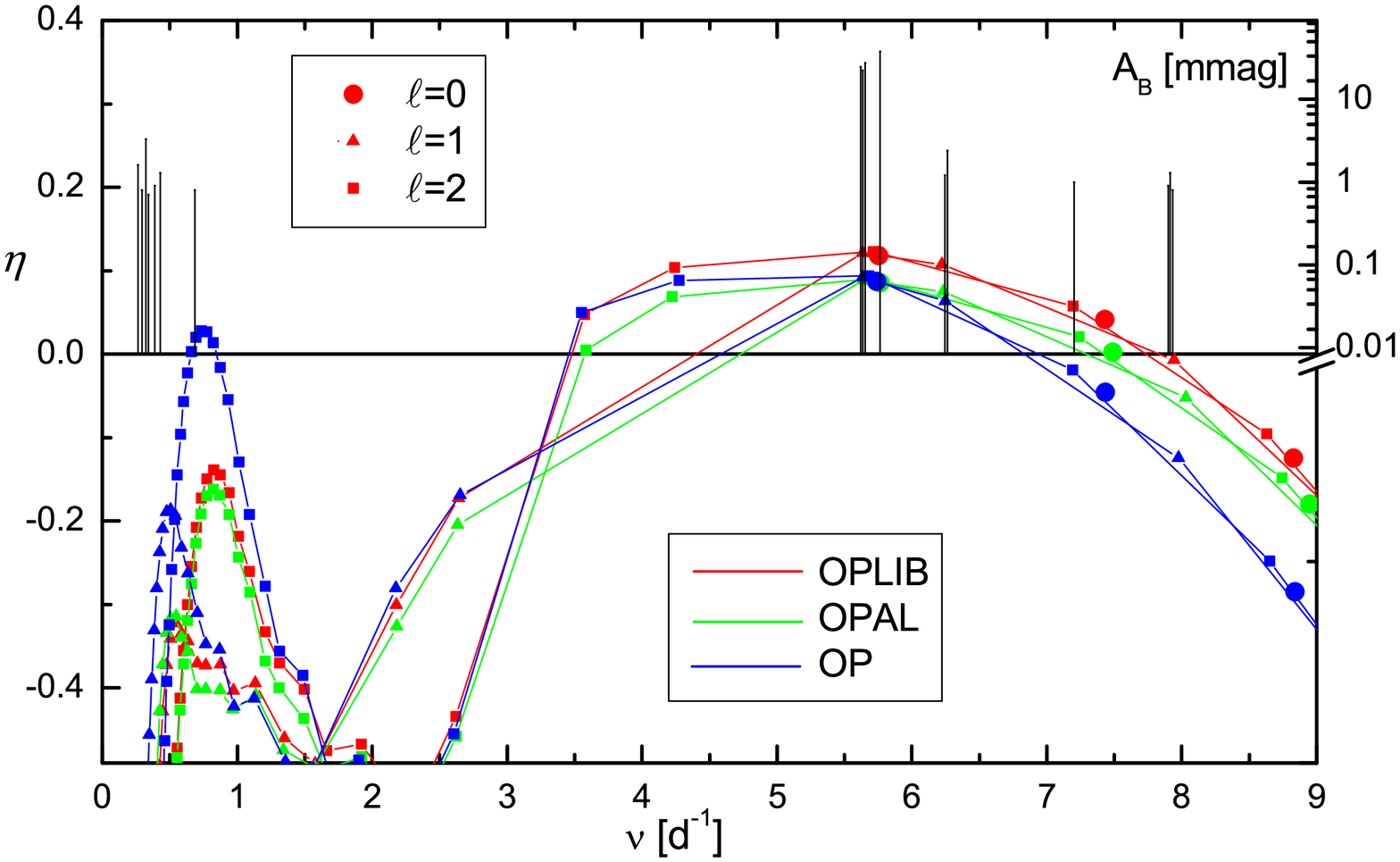}
\includegraphics[clip,width=0.48\linewidth]{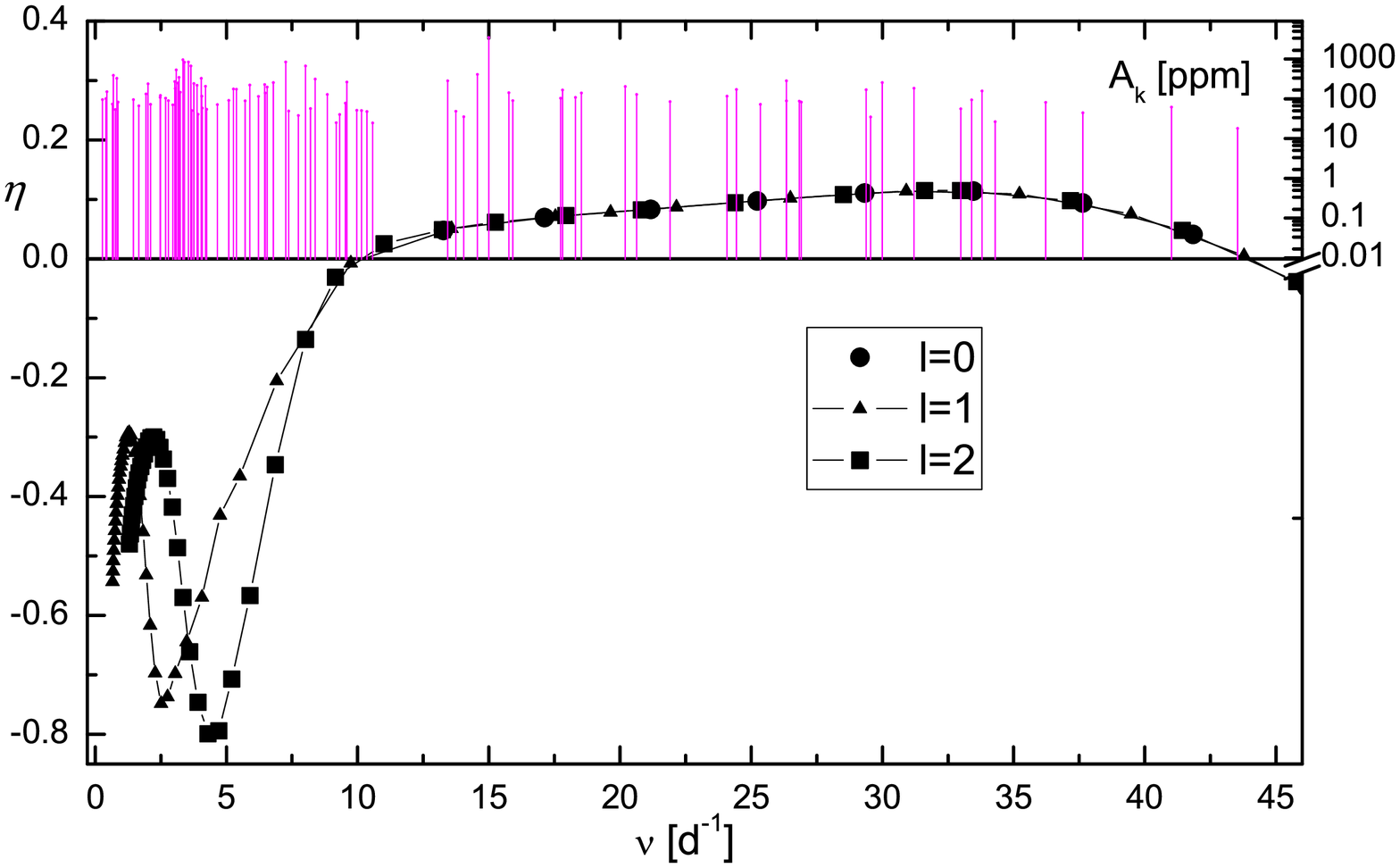}
	\caption{{\it The left panel}: The normalized instability parameter, $\eta$, as a function the mode frequency for representative seismic models of $\nu$ Eridani, computed with
the three opacity data: OPLIB, OPAL and OP. All models have  $M=9.5~M_\odot$, $\log T_{\rm eff}\approx 4.343$, $Z=0.015$ and the overshooting parameter $\alpha_{\rm ov}\approx 0.08$.
{\it The right panel}: A similar plot but for the $\delta$ Sct model with $M=1.8~M_\odot$, $\log T_{\rm eff}\approx 3.887$ computed with the OPAL tables.
The values of the Kepler amplitudes of KIC\,8197788 are on the right Y-axis.}
	\label{fig2}
\end{figure*}

\section{Constraints on opacities}

The main goal of our seismic modelling of  $\beta$ Cep/SPB and $\delta$ Sct stars was to get instability in the whole range of the observed frequencies.
To this aim we modified the standard opacity tables by increasing or reducing the mean opacity in the temperature range $\log T\in(5.0,~5.5)$
with the step $\Delta\log T_0=0.005$. For more details see \citet{JDD2017}.

To limit the  number of possible solutions we fitted also the parameter which describes the relative amplitude of the radiative flux perturbations at the photosphere level.
This amplitude is the so-called nonadiabatic parameter $f$ and its
empirical values can be derived from multi-colour light variations and radial velocity measurements \citep[e.g.,][]{JDD2003}.

Up to now, we performed such instability analysis for  six early B-type pulsators  dominated by p modes: $\nu$ Eri, 12 Lac, $\gamma$ Peg, $\theta$ Oph, $\kappa$ Sco and $\alpha$ Lup,
one  early B-type pulsator dominated by g modes (KIC\,3240411) and one late B-type pulsator dominated by g modes (KIC\,11971405).
B-type main sequence stars are relatively simple objects because there is no transport by convection in their envelopes and no significant mass loss occurs.
In Table\,1, we list the seismic models with the opacity modifications that were indispensable to account for the whole oscillation spectra of the studied stars.
As one can see each B-type pulsator demands a high increase of opacity near $\log T=5.46$ which corresponds to the maximum contribution of nickel.
This ''nickel'' opacity increase is necessary to excite high-order g modes. The same solution was proposed by \citet{Salmon2012}
who analysed pulsations in B-type stars of the Magellanic Clouds.
\begin{table}[h]
	\caption{The opacity modifications obtained from seismic studies of B-type pulsators. }
	\label{table1}
	\begin{tabular*}{1.1\linewidth}{l c c c c c c c c}
	\noalign{\smallskip}\hline\hline\noalign{\smallskip}
  star & $M/M_\odot$ &  $\log (T_{\rm eff})$ & $\log T_{0,1}$ & $\Delta\kappa$ & $\log T_{0,2}$ & $\Delta\kappa$ & $\log T_{0,3}$ & $\Delta\kappa$  \\
       &             &                     &                &    [\%]        &                &    [\%]        &                &    [\%]         \\
	\noalign{\smallskip}\hline\noalign{\smallskip}
	$\nu$ Eridani &  9.0 &  4.331 & 5.06 & $-60$ &  5.22  &  $+35$ & 5.46 & $+220$ \\
& & & & & & \\
	 12 Lacertae  & 11.2 &  4.376 & 5.06 & $-25$ &  5.22  &  $+50$ & 5.46 & $+200$ \\
& & & & & & \\
  $\gamma$ Pegasi & 8.1 & 4.324 & 5.06 & $-60$ &  5.22  &  $+50$ & 5.46 & $+210$ \\
  & & & & & & \\
$\theta$ Ophuichi & 8.4 & 4.343 & 5.06 & $+30$ &  5.30  &  $+65$ & 5.46 & $+145$ \\
& & & & & & \\
 $\kappa$ Scorpii & 10.4 & 4.363 & 5.06 & $+30$ &  5.22  &  $+30$ & 5.46 & $+100$ \\
 & & & & & & \\
$\alpha$ Lupi & 12.0  & 4.351 & -- & -- &  --  &  -- & 5.46 & $+100$ \\
 & & & & & & \\
 KIC3240411 & 6.35 & 4.294 & 5.06 & $-50$ &  5.22  &  $+50$ & 5.46 & $+200$ \\
 & & & & & & \\
 KIC11971405 & 4.55 & 4.167 & 5.06 & $-50$ &  5.22  &  $+50$ & 5.46 & $+200$ \\
	\noalign{\smallskip}\hline
	\end{tabular*}
\end{table}

An attempt to account for all frequencies detected in $\delta$ Scuti stars has been made by \citet{Balona2015}.
For all $\delta$ Sct stars, periodograms derived from the Kepler photometry have shown rich low-frequency spectra which
can correspond only to high-order gravity modes \citep{Balona2014}.
These stars have effective temperatures higher than the granulation boundary where the depth of the convective zone is too thin
to drive $\gamma$ Doradus pulsations. \citet{Balona2015} have shown that increasing opacity
at $\log T=5.06$ by a factor of two to three allowed
to excite dipole and quadrupole low-frequency modes in the corresponding pulsational models.
Any further increase of opacity resulted in saturation and no further increase in instability.

\section{Summary}

We briefly recapped our most important results on complex seismic analysis of main sequence pulsators in order to obtain constraints on stellar opacities.
There are two main messages from pulsational analysis of hybrid main sequence pulsators. Firstly, there is a problem explaining their whole
oscillation spectra using standard opacity models. Secondly, the likely reasons for the discrepancy between theory and observations
are still existing uncertainties in opacity data. To excite modes corresponding to all observed frequencies,
huge opacity modifications  are needed at some temperatures.

However, to draw plausible conclusions, it is very important to control a number and a type of opacity modifications.
To this end, we add the requirement to reproduce the relative amplitude of the bolometric flux variations, i.e., the parameter $f$.
This requirement significantly limits the number of solutions because the parameter $f$
is very sensitive to the structure of subphotospheric layers where the pulsation driving occurs.
Such complex seismic studies have to be performed for more stars.

\acknowledgements{The work was financially supported by the Polish NCN grants 2018/29/B/ST9/02803 and 2018/29/B/ST9/01940.}

\bibliographystyle{ptapap}
\bibliography{pta_JDD}

\end{document}